\documentclass{article}
\usepackage{amsmath}
\usepackage{amssymb}
\usepackage{graphicx}

\begin{document}

\title{Strong necessary conditions and Cauchy problem}  

\author{{\L}ukasz T. St\c{e}pie\'{n} \thanks{Institute of Computer Science, 
The Pedagogical University of Cracow, 
ul. Podchorazych 2, 30-084 Krak\'{o}w, Poland, e-mail: sfstepie@cyf-kr.edu.pl}}


\maketitle

\begin{abstract}
Some exact solutions of boundary or initial conditions formulated for Bogomolny equations (derived by using the strong necessary conditions and associated with some ordinary equation and some partial differential equations), have been found. Besides, a degeneracy of the hamiltonian for the restricted baby Skyrme model has been established. 
\end{abstract}

keywords: action integral, Bogomolny equations, Bogomol'nyi equations, boundary and initial conditions, Cauchy problem, degeneracy of  hamiltonian, degeneracy of energy, energy degeneracy, strong necessary conditions\\ 

\section{Introduction}\label{I}
There are several approaches to solving of nonlinear partial differential equations (see for e.g.:  
\cite{Adomian1994}, \cite{Atmaja2017},
\cite{Atmaja2018}, \cite{Benci},
\cite{Bluman}, \cite{Cieslinski}, \cite{Conte}, \cite{Debnath}, \cite{Doliwa}, \cite{Elzaki}, \cite{Feng}, \cite{Fushchich1993},
\cite{Gaeta}, \cite{Goldstein}, \cite{Gu}, \cite{Kudryashov}, \cite{Magri}, \cite{Meleshko2005}, \cite{Musette}, 
\cite{Krasilshchik},  \cite{Nieszporski},  
\cite{Olver}, \cite{PolyaninZaitsev}, \cite{Rajchel}, \cite{Rubina}, \cite{Satsuma}, \cite{ST2010}, \cite{Vitanov}, \cite{WIET2005}, \cite{Winternitz},  \cite{Zakharov1979}, \cite{Zhang}, and proper references therein).
Four decades ago another method - the so-called the strong necessary conditions method (SNCM) was formulated for solving nonlinear partial differential equations (resulting from variational principles). We were interested in exact analytic solutions. Main idea leading to achievement was to replace the Euler-Lagrange equations by other variational method, which possessed the order smaller then the original ones. Moreover, the set of the solutions derived by the considered method, has to be included in the set of the solutions of the original Euler-Lagrange equations. 
 Crucial role in SNCM is played by the set of topological invariants. The set of solutions of NPDE depends on the subset of implemented invariants. The empty subset of invariants always corresponds to empty set or set of trivial solutions. For some simple examples of the  applications of SNCM, see references \cite{Adam_Santamaria}, \cite{SOK1}, \cite{SOK1981}, \cite{SOK1a}, 
\cite{SokStSok}, \cite{SOK2}, \cite{SOK2a}, 
\cite{SOK22}, \cite{SOK3}, \cite{ST1}, \cite{SOK4}, \cite{ST2}, \cite{ST2015}, \cite{ST11}, \cite{ST12}, \cite{ST3}. In 2001 Professor Boles{\l}aw Szafirski has pointed out that it is unknown, how to implement boundary and initial conditions as well as how to set the Cauchy problem in SNCM,  \cite{Szafirski}. This paper shows a way for satisfying His requirement. 
In the section 2 we present briefly the concept of strong necessary conditions. The next section is devoted to solving Cauchy problem
associated with Bogomolny equation (which is an ordinary differential equation) of 1-dimensional harmonic oscillator. 
In the section 4 we present solutions of Cauchy problem associated with Bogomolny equations for continous Heisenberg model and restricted baby Skyrme model. The section 5 includes some conclusions. 

\section{A short presentation of the method}
 The main idea of the concept of strong necessary conditions \cite{SOK1} - \cite{SOK3}, is such that instead of considering the Euler-Lagrange equations, 
 
 \begin{equation}
 F_{,u} - \frac{d}{dx}F_{,u_{,x}} - \frac{d}{dy}F_{,u_{,y}}=0, \label{el}
 \end{equation}
  which follow from the varying of the functional
 
 \begin{equation}
 \Phi[u]=\int_{E^{2}} F(u,u_{,x},u_{,y}) \hspace{0.05 in} dxdy, \label{functional}
 \end{equation}

one considers strong necessary conditions, \cite{SOK1}, \cite{SOK1981}, \cite{SOK1a}, \cite{SOK2}, \cite{SOK2a}, \cite{SOK22}, 
\cite{SOK3}:

 \begin{gather}
   F_{,u}=0, \label{silne1} \\
   F_{,u_{,x}}=0, \label{silne2} \\
   F_{,u_{,y}}=0, \label{silne3}
 \end{gather} 

 where $F_{,u} \equiv \frac{\partial F}{\partial u}$, etc.
   
 Obviously, the set of the solutions of the system of the equations (\ref{silne1}) - (\ref{silne3}), is a subset of the set of the solutions satisfying the Euler-Lagrange equation (\ref{el}). On the other hand, even if this subset is non-empty, its elements (solutions of the system (\ref{silne1}) - (\ref{silne3})), are very often trivial solutions. So, in order to extend this subset, we consider the 
gauged functional (\ref{functional})
 
  \begin{equation}
  \tilde{\Phi} = \Phi + I, \label{gauge_transf}
  \end{equation}
  where $I$ is such functional that its local variation vanishes, with respect to $u(x,y)$:
 $\delta I \equiv 0$.
 
  Owing to this feature, the Euler-Lagrange equations (\ref{el}) possess the same form as the Euler-Lagrange equations resulting from  requiring of the extremum of (\ref{gauge_transf}). 
 On the other hand, the equations following from applying of the strong necessary conditions (\ref{silne1}) - (\ref{silne3}), to 
(\ref{gauge_transf}), do not possess the same form as the equations following from applying the strong necessary conditions (\ref{silne1}) - (\ref{silne3}), to (\ref{functional}). Hence, there is an opportunity to obtain non-trivial solutions. Let us note that the order of the system  of the partial differential equations, constituted by strong necessary conditions (\ref{silne1}) - (\ref{silne3}), is less than the order of Euler-Lagrange equations (\ref{el}). The method of derivation Bogomolny equations (Bogomolny decomposition), by using the strong necessary conditions, was included and applied in \cite{SOK1}, \cite{SokStSok}, \cite{ST1}, and developed in 
\cite{Adam_Santamaria}. As we see, this approach differs from the approach of deriving of Bogomolny equations, presented in 
\cite{BelPol}, \cite{BPST}, \cite{Bialynicki}, \cite{Bogomolny}, \cite{Hosoya}. In \cite{ST2}, the Bogomolny equations for baby Skyrme models, were derived, by using the concept of strong necessary conditions. The idea that the lagrangian after adding to it a total derivative of a function dependent only on field variable, generates the same Euler-Lagrange equations as the original lagrangian, had been known very well in the literature ({\em{c.f.}} for e.g. \cite{ArodzHadasz}, \cite{Bogoljubow}). However, gauging the lagrangian on a complete set of invariants and applying this to derivation of Bogomolny equations, by using the concept of strong necessary conditions, 
was firstly presented just in \cite{SOK1}.

\section{Ordinary Differential Equations}\label{II}
In this section we present application of SNCM in an initial conditions problem.

As an introductory example we consider linear equation resulting from the SNCM applied to Lagrangian of the one dimensional harmonic  oscillator:
\begin{equation}
\label{Ia}
\mathcal{L}=\frac{m}{2}\bigg(\bigg(\frac{dx}{dt}\bigg)^2-\omega^2x^2\bigg).
\end{equation}

In order to set the strong necessary conditions we perform the gauge transformation of (\ref{Ia}) using the following topological invariant density $G(x) \frac{dx}{dt}$, where $G(x)$ is an arbitrary function and $G(x) \in \mathcal{C}^{1}$:

\begin{equation}
\label{IIa}
\tilde{\mathcal{L}}=\frac{m}{2}\bigg(\bigg(\frac{dx}{dt}\bigg)^2-\omega^2 x^2\bigg) + G(x) \frac{dx}{dt}.
\end{equation}

Note that $\mathcal{L}$ depends on the two functions:$\mathcal{L}=\mathcal{L}(x,\frac{dx}{dt})$. According to the strong necessary conditions we have to optimize the action functional, by regarding both, $x$ and $\frac{dx}{dt}$:

\begin{equation}
\label{IIIc}
\frac{\partial \tilde{\mathcal{L}} }{\partial x}=0,\hspace{3mm} \frac{\partial \tilde{\mathcal{L}} }{\partial (\frac{dx}{dt})}=0.
\end{equation}

Equations (\ref{IIIc}) read:

\begin{equation}
-m\omega^2x+G_{x} \frac{dx}{dt}=0,\label{IV}\end{equation}
\begin{equation}G+m\frac{dx}{dt}=0.\label{V}
\end{equation}

We eliminate $\frac{dx}{dt}$, from this system, and we get the equation, which has to be satisfied by the function $G$: 

\begin{equation}
G G_{,x} + m^{2} \omega^{2} x = 0.
\end{equation}

Hence

\begin{equation}
\frac{1}{2} (G^{2})_{,x} + m^{2} \omega^{2} x = 0.
\end{equation}

The solution of this equation has the form

\begin{gather}
G = \pm \sqrt{c_{1} - m^{2} \omega^{2} x^{2}}.  \label{rozwG}
 \end{gather} 

Then, we formulate Cauchy problem
 
 \begin{gather}
-m\omega^2x+G_{x} \frac{dx}{dt}=0,\label{Cauchy1} \\
G+m\frac{dx}{dt}=0,\label{Cauchy2} \\
 x(0) = c_{3}. \label{Cauchy3}
\end{gather}

Solving (\ref{Cauchy1}) - (\ref{Cauchy2}), provided that (\ref{rozwG}), where we take into account "plus" sign, we get

\begin{equation}
\label{rozw_Cauchyjednow}
x(t) = \frac{ \sqrt{c_{1}}
 \tan{(\omega (c_{2} - t))}}{\omega m \sqrt{(\tan^{2}{(\omega (c_{2}- t))} + 1)} }, 
\end{equation}

where $c_{2}$ is the integration constant. Now we take into account (\ref{Cauchy3}), hence

\begin{equation}
c_{2}=\frac{1}{\omega} \arctan{\frac{m \omega c_{3}}{\sqrt{c_1 - c^{2}_{3} \omega^{2} m^{2}}}} \label{c2}
\end{equation}

If we take into account the Euler-Lagrange equations for this problem

\begin{equation}
m \frac{d^{2} x(t)}{dt^{2}} + m \omega^{2} x(t) = 0, \label{EulerLagrange}
\end{equation}

then its solution is

\begin{equation}
  x(t) = A \sin{(\omega t)} + B \cos{(\omega t)},
 \end{equation}

 where $A=const, B=const$, and this does not satisfy the Bogomolny equations (\ref{Cauchy1}) - (\ref{Cauchy2}), where $G$ is given by (\ref{rozwG}). Obviously, the solution of Bogomolny equations, given by (\ref{rozw_Cauchyjednow}), with and without providing that
 (\ref{c2}), satisfies (\ref{EulerLagrange}).

\section{Partial Differential Equations}

\subsection {Field Equations and Cauchy problem associated with $\pi_2(S^2)$ homotopy group}

As an example we consider the continous Heisenberg model represented by the following Hamiltonian, \cite{BelPol}: 
\begin{equation}
\label{9}
\mathrm{H}=\int_{E^2} \bigg (\frac{\nabla w\cdot\nabla w^*}{(1+w\cdot w^*)^2}+I_1 \bigg)dxdy,
\end{equation}
where the field variable $w$ consists of classical spin components:
 \begin{equation}
  \label{10}
 w=\frac{(S^x+iS^y)}{(1+S^z)}
 \end{equation}
 where $S^\alpha$ are components of the classical spin. $I_{1}$ is density of the  topological invariant:
 \begin{equation}
 \label{11}
 I_{1}=G_{1}(w,w^{*})(w_{,x} w^{*}_{,y}-w_{,y} w^{*}_{,x}),
 \end{equation} 
 
 We apply the strong necessary conditions to (\ref{9}) and we obtain the system of dual equations, which can be also obtained
 as a two-dimensional version of the system of the dual equations derived in \cite{SokStSok}:
 
 \begin{gather}
 -\frac{2 w^{\ast} \nabla w \nabla w^{*}}{(1+ w w^{*})^{3}} + G_{1, w}(w_{,x} w^{*}_{,y}-w_{,y} w^{*}_{,x})
  +  D_{x} G_{1, w}(w, w^{\ast}) + D_{y} G_{2, w}(w, w^{\ast}) = 0, \\
	c.c.,\\
	\frac{w^{*}_{,x}}{(1+w w^{*})^{2}} + G_{1} w^{*}_{,y} + G_{2, w} = 0, \label{dolne1}\\
	\frac{w^{*}_{,y}}{(1+w w^{*})^{2}} - G_{1} w^{*}_{,x} + G_{3, w} = 0, \label{dolne2} \\
  c.c. \label{dolne3}
 \end{gather}  

  We make this system self-consistent by choosing $G_{n} = const$ ($n = 2, 3$) and (as in \cite{SokStSok} by choosing
 $G_{1} = \frac{i}{(1+ w w^{*})^{2}}$. 
 Next, expressing the complex fields $w$ and $w^*$ 
 by real fields:
 \begin{equation}
 	\label{12}
 w=U(x,y) + iV(x,y), w^*= U(x,y) - iV(x,y),
 \end{equation}
  
we derive from (\ref{dolne1}) - (\ref{dolne3}), the pair of equations, governing real fields $V(x,y)$ and $U(x,y)$:
\begin{equation}
\label{CR1} 
{\frac {\partial }{\partial y}}V \left( x,y \right) -{\frac {\partial }{\partial x}}U \left( x,y \right) =0.
\end{equation}	
\begin{equation}
\label{CR2}
	{\frac {\partial }{	\partial x}}V \left( x,y \right) +{\frac {\partial }{\partial y}}U\left( x,y \right) =0
\end{equation}

Solving (\ref{CR1}) and(\ref{CR2}) we get:

\begin{equation}
\label{15}
U( x,y) =F_1( y-ix ) + F_2( y+ix ),
\end{equation}
\begin{equation}
\label{16}
V( x,y ) =-iF_1( y-ix ) +iF_2( y+ix) +C_1,
\end{equation}
where $F_1(\cdot)$ and $F_2(\cdot)$ are some functions. After taking into account the formula (\ref{12}), we obtain that $F_{1}, F_{2}$ are connected with $w, w^{\ast}$, by the formulas

\begin{gather}
F_{1} = \frac{1}{2}(w - iC_{1}),\\
F_{2} = \frac{1}{2}(w^{\ast} + iC_{1})
\end{gather}

and $C_{1}$ is an arbitrary real constant. \\

  Basing on the general solutions (\ref{15}),(\ref{16}) of (\ref{CR1}), (\ref{CR2}) we present the Cauchy problem for partial differential equations of the first order created by the strong necessery conditions.
The considered example consists of two independent variables $x$ and $y$ and two functions. Therefore it is possible to formulate the following constrains for the general solutions:
\begin{equation}
\label{17}
 U(x,0)=f_1(x),\hspace{2mm}V(x,0)=f_2(x),
\end{equation} 
where $f_1(x)$ and $f_2(x)$ are given functions.\\ 
It is possible for the considered Heisenberg model to derive analogous relations to
$U(0,y)$ and $V(0,y)$, which relate integration constants to initial or boundary conditions.
 Constraining (\ref{15}) and (\ref{16}) to (\ref{17}) and substituting $y=0$ we obtain:
\begin{equation}
\label{18}
f_1(x)=F_1(-ix)+F_2(ix),
\end{equation}
\begin{equation}
\label{19}
f_2(x)=-iF_1(-ix)+iF_2(ix)+C_1,
\end{equation}
Since $f_1(x)$ and $f_2(x)$ are given therefore $F_1$ and $F_2$ can't be arbitrary:
\begin{equation}
F_1(-ix)=i{ f_2(x)}+{ f_1(x)}/2+{ f_2(x)}/2-i/2\hspace{1mm} C_1
\end{equation}
\begin{equation}
F_2(-ix)=\frac{if_1(x)+f_2(x)-C_1}{2i}.
\end{equation}
Therefore, the only freedom for $F_1$ and $F_2$ is gauge transformation regarding $C_1$ constant. This full solution can be extended
by applying semi-strong necessary conditions concept (this concept was presented in \cite{SOK2}).

\section{Field Equations and Cauchy problem for the restricted baby Skyrme model}

 The restricted baby Skyrme model has the following hamiltonian

 \begin{equation}
   \mathcal{H} = -4\beta\frac{(\omega_{,x}\omega^{\ast}_{,y}-\omega_{,y}\omega^{\ast}_{,x})^{2}}{(1+\omega 
   \omega^{\ast})^{4}} + V(\omega,\omega^{\ast}), \label{hamilt}
   \end{equation}
	
	In \cite{ST2}, the Bogomolny decomposition for this model, was derived by using the concept of strong necessary conditions 
		(the Bogomolny equations for this model, but for some special forms of the potential, and by another way, and some solutions of these equations, were derived in \cite{AdamEtAl2}). 
	We apply this concept to the hamiltonian gauged on the invariants is as follows, \cite{ST2}
	
	\begin{equation}
	\tilde{\mathcal{H}} = -4\beta\frac{(\omega_{,x}\omega^{\ast}_{,y}-\omega_{,y}\omega^{\ast}_{,x})^{2}}{(1+\omega 
   \omega^{\ast})^{4}} + V(\omega,\omega^{\ast}) + G_{1} (\omega_{,x}\omega^{\ast}_{,y}-\omega_{,y}\omega^{\ast}_{,x}) +
	 D_{x} G_{2} + D_{y} G_{3}, \label{hamilt_przecech}
	 \end{equation}
	
	where $G_{i}, (i = 1, 2, 3)$ are some unspecified functions of $\omega, \omega^{\ast}$ (of course, $G_{i} \in \mathcal{C}$).  
	If 
	
	\begin{gather}
	G_{1} = \frac{4i\sqrt{\beta}}{(1+\omega \omega^{\ast})^{2}} \sqrt{V(\omega, \omega^{\ast})}, \ 
	G_{k} = const, (k = 2, 3), \label{warunki_na_G}
	\end{gather}
	
	we can derive the Bogomolny decomposition, which in this case, has the following form, \cite{ST2}

\begin{gather}
  \omega_{,x}\omega^{\ast}_{,y}-\omega_{,y}\omega^{\ast}_{,x} = \frac{i}{2\sqrt{\beta}} \sqrt{V(\omega, \omega^{\ast})} 
  (1+\omega\omega^{\ast})^{2}. 
  \label{bogomolny_decomp}
   \end{gather}

	We find now an exact localized solution (with localised density of energy), of the Bogomolny decomposition (\ref{bogomolny_decomp}), for the case of the so-called, "Mexican hat" potential: $V=\lambda_{3}(\omega \omega^{\ast} - \gamma^{2})^{2}$. We use "hedgehog ansatz":
	
	 \begin{equation}
  \omega=\frac{\sin{(f(r))} \cos{(N\theta)} + i \sin{(f(r))} \sin{(N\theta)}}{1+\cos{(f(r))}}, 
  c.c.,
  \end{equation}
	
		where $(r, \theta)$ are polar coordinates in the cartesian $x - y$ plane.  
	
	After inserting this ansatz into (\ref{bogomolny_decomp}), we formulate the Cauchy problem

\begin{gather}
\frac{ (\cos{(f(r))} + 1) N f'(r) \sin{(f(r))}}{r} = \sqrt{\frac{\lambda_{3}}{\beta}} \bigg[\cos{(f(r))} (\gamma^{2} + 1) + \gamma^{2} - 1 \bigg],\\
  f(0) = c_{0}= const, 
\end{gather}

where in this case we put $c_{0}=2$.

 We are interested in obtaining some localized solution, then we impose also the conditions:

 \begin{gather} 	
	\lim_{r\rightarrow \pm \infty} f(r) = const,\\
  \lim_{r\rightarrow \pm \infty} \mathcal{H} = const, 
 \end{gather} 

We solve this problem and we have 

\begin{eqnarray}
\begin{gathered}
   f(r) = \pi - \arccos{\bigg( X_{1} \bigg) },
 \end{gathered}
 \end{eqnarray}

 where 

 \begin{gather}
 X_{1} = \frac{1}{\gamma^{2} + 1} \bigg(\gamma^{2} - \\ 
	exp\bigg(\frac{1}{\sqrt{\beta} N} \bigg( -4 Lambert \bigg( \frac{1}{2} exp\bigg(\frac{1}{N\sqrt{\beta}} X_{2}\bigg) \bigg) \times \\
						N\sqrt{\beta} + X_{2} \bigg) \bigg) - 1 \bigg) 
						\end{gather} 

 \begin{gather}
 X_{2} =  N \ln{((\gamma^{2} + 1) \cos{(2)} + \gamma^{2} - 1)} \sqrt{\beta} + \\
\frac{N (\gamma^{2} + 1) (\cos{(2)} + \gamma^{2} - 1) \sqrt{\beta}}{2} - 
\frac{\sqrt{\lambda_{3}} r^{2} (\gamma^{2} + 1)^{2}}{4},
 \end{gather}
 		
						$Lambert(Y)$ is the so-called Lambert function, which satisfies the equation $Lambert(Y) \exp{(Lambert(Y))} = Y$.

For $\gamma = 2, N = 1, \lambda_{3} = 1, \beta = 1$:

 \begin{equation}
 \begin{gathered}
  f(r)= \arccos{\bigg\{\bigg[\frac{2 Lambert\bigg(\frac{1}{2} 
	\exp{\bigg(-\frac{25}{4} r^{2} + \frac{5\cos{(2)} +3}{2}\bigg)} (5\cos{(2)}+3) \bigg) -3}{5} \bigg]\bigg\}}. \label{rozw_na_f}
  \end{gathered}
  \end{equation}
	
	We present a figure of this above solution on FIG 1.
	
\begin{figure}[htb]
\centerline{%
\includegraphics[width=12.5cm]{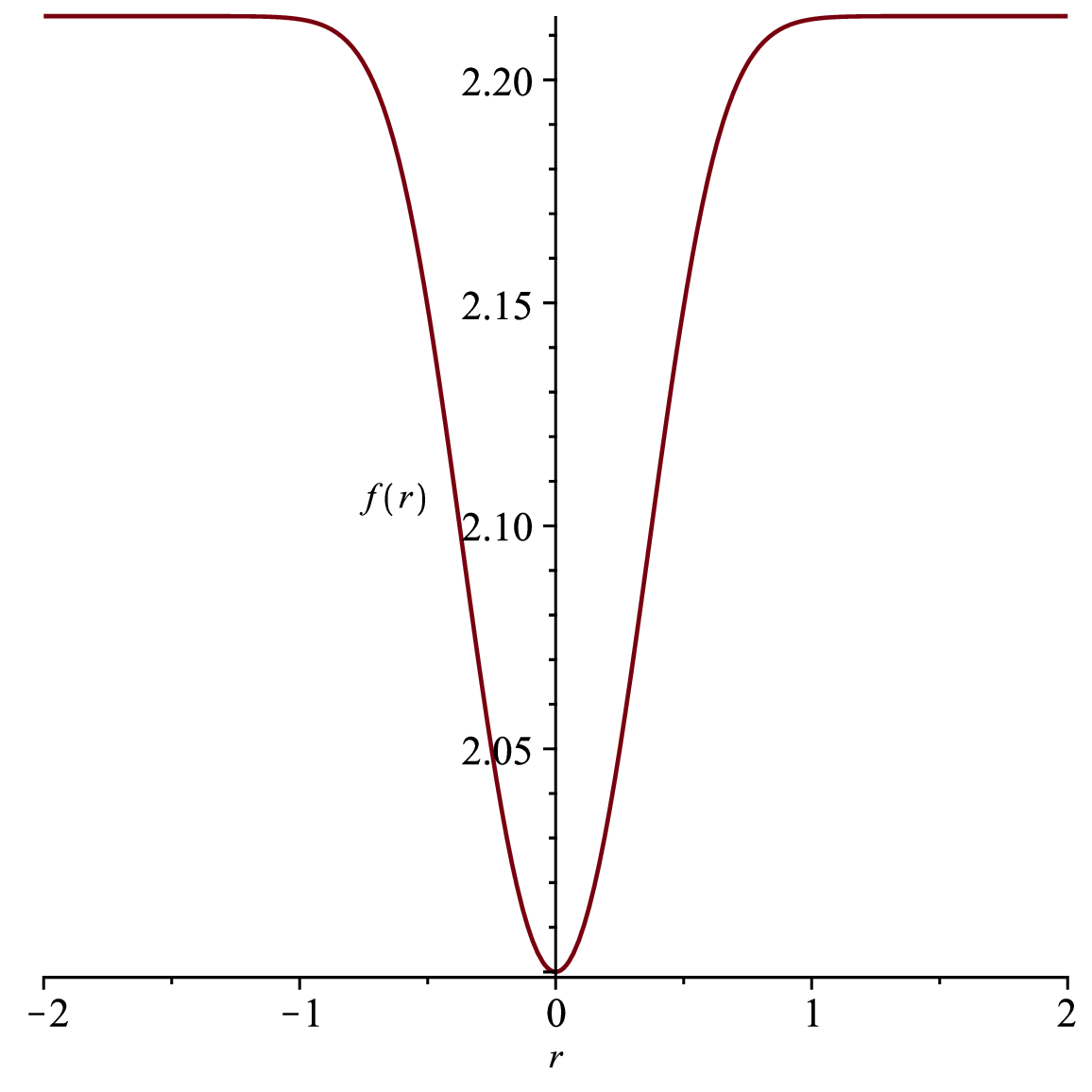}}
\caption{Plot of the solution (\ref{rozw_na_f})}
\label{Fig:F2H}
\end{figure}

Now, it has turned out that if we insert the found solution of Cauchy problem into the ungauged and gauged hamiltonian densities 
(\ref{hamilt}), (\ref{hamilt_przecech}), correspondingly, then ungauged hamiltonian density is nonzero and its figure is on FIG 2.

\begin{figure}[htb]
\centerline{%
\includegraphics[width=12.5cm]{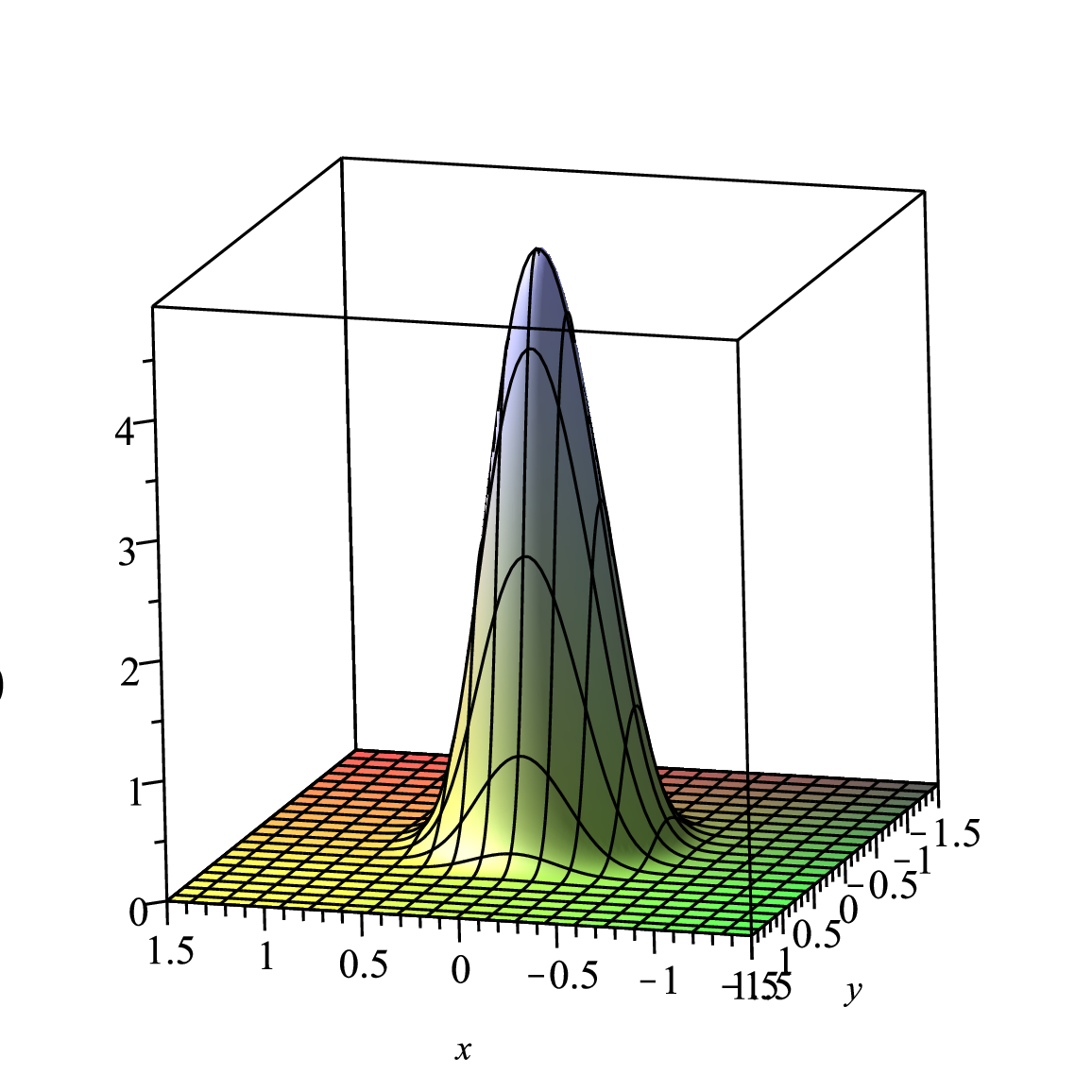}}
\caption{Plot of the ungaued hamiltonian density for the solution (\ref{rozw_na_f})}
\label{Fig:F3H}
\end{figure}

The gauged hamiltonian density is zero (of course, the condition (\ref{warunki_na_G}) and Bogomolny equations (\ref{bogomolny_decomp})
hold): 

\begin{equation}
\tilde{\mathcal{H}} = 0.
\end{equation}

Thus, we can tell here about a degenerate hamiltonian, (the problem of degenerate hamiltonian in the case of theory of gravity, was  investigated in \cite{Sanyal}; there in \cite{Castillo} was proven the existence of an infinite number of Lagrangians for a given second-order ODE). It corresponds to the fact that if we consider two versions of a field-theoretical lagrangian: ungauged and gauged on total derivatives of any function of field variables, then energy-momentum tensors corresponding to each of these lagrangians, will be different, \cite{Arodz}. 

\section{Conclusions}
The first conclusion concerns just possibility to solve the ordinary differential equations subjected to the strong necessary conditions. In the case of linear ODE, the solution of Cauchy problem is given by (\ref{rozw_Cauchyjednow}) - (\ref{c2}). \\
The formulas (\ref{CR1}) and (\ref{CR2}) establish Cauchy-Riemann system, which is a start point for the theory of analytic functions. Hence, because of Riemann theorem, this may be a step to the investigations of conformal maps.  \\
Moreover, as far as the Cauchy problems for Heisenberg model and for restricted baby Skyrme model, are concerned, after using of strong necessary conditions and deriving Bogomolny equation for this problem, one can formulate Cauchy problem and solve it. \\
We have also obtained some localized solution of Cauchy problem of Bogomolny equation for restricted baby Skyrme model. 
An example of such solution is given by (\ref{rozw_na_f}). 
Besides, we have also showed on the example of restricted baby Skyrme model, that there exists degeneracy of hamiltonian, i.e. the values
of hamiltonians: ungauged and gauged one, are different for the solution of Cauchy problem for Bogomolny equations, and these both  hamiltonians generate the same Euler-Lagrange equations.  

\section{Acknowledgments}
The author thanks to Prof. K. Sokalski for very fruitful discussions. 
Some part of computations was carried out by using Maple Waterloo Software, on the computer "mars"  (grant nr MNiSW/IBM$\_$BC$\_$HS21/AP/057/2008) in ACK-CYFRONET AGH in Cracow, and owing to a financial support, provided by The Pedagogical University of Cracow, within a research project (the leader of this project: Dr K. Rajchel). This research was supported partially also by PL-Grid Infrastructure.

 \bibliographystyle{plain} 
 \bibliography{L_T_Stepien_bibtech}

\providecommand{\noopsort}[1]{}\providecommand{\singleletter}[1]{#1}%
\begin{thebibliography}{10}

\bibitem{AdamEtAl2}
C.~Adam, T.~Roma\'{n}czukiewicz, J.~Sanchez-Guillen, and A.~Wereszczy\'{n}ski.
\newblock {\em Phys. Rev.}, {\bf{D81}}:085007, 2010.
\newblock hep-th/1002.0851.

\bibitem{Adam_Santamaria}
C.~Adam and F.~Santamaria.
\newblock {\em J. High Energ. Phys.}, {\bf{2016}}:047, 2016.
\newblock arXiv:1609.02154.

\bibitem{Adomian1994}
G.~Adomian.
\newblock {\em Solving Frontiers Problems of Physics: The Decomposition
  Method}.
\newblock Springer Science+Business Media, Dordrecht, 1994.

\bibitem{Arodz}
H.~Arod\'{z}.
\newblock Lectures on field theory.
\newblock Delivered at Institute of Physics, Jagiellonian University 1997/1998
  (unpublished).

\bibitem{ArodzHadasz}
H.~Arod\'{z} and L.~Hadasz.
\newblock {\em Lectures on Classical and Quantum Theory of Fields}.
\newblock Springer-Verlag Berlin Heidelberg, 2017.

\bibitem{Atmaja2017}
A.~N. Atmaja.
\newblock {\em Phys. Lett.}, {\bf{ B768}}:351, 2017.

\bibitem{Atmaja2018}
A.~N. Atmaja.
\newblock hep-th/1807.01483, 2018.

\bibitem{BelPol}
A.~A. Belavin and A.~M. Polyakov.
\newblock {\em JETP Lett.}, {\bf{22}}:245, 1975.

\bibitem{BPST}
A.~A. Belavin, A.~M. Polyakov, A.~S. Schwartz, and Yu.~S. Tyupkin.
\newblock {\em Phys. Lett.}, {\bf{B59}}:85, 1975.

\bibitem{Benci}
V.~Benci and D.~Fortunato.
\newblock {\em Variational Methods in Nonlinear Field Equations. Solitary
  Waves, Hylomorphic Solutions and Vortices}.
\newblock Springer International Publishing Switzerland, 2014.

\bibitem{Bialynicki}
I.~Bia{\l}ynicki-Birula.
\newblock On the stability of solitons.
\newblock In Antonio~F. Ra$\tilde{n}$ada, editor, {\em Nonlinear Problems in
  Theoretical Physcis, Jaca, Huesca (Spain), June , 1978}, volume~98 of {\em
  Lect. Notes Phys.}, page~15. Birkh\"{a}user, Springer Basel, 1979.

\bibitem{Bluman}
G.~W. Bluman, A.~F. Cheviakov, and S.~C. Anco.
\newblock {\em Applications of Symmetry Methods to Partial Differential
  Equations}.
\newblock Springer Science+Business Media, 2010.

\bibitem{Bogoljubow}
N.~N. Bogoliubov and D.~V. Shirkov.
\newblock {\em Introduction to the Theory of Quantized Fields}, volume~10 of
  {\em Nikolai Nikolaevich Bogoliubov. Collection of Scientific Works}.
\newblock Moscow Nauka, 2008.
\newblock (in Russian), Editor-in-Chief A. D. Sukhanov.

\bibitem{Bogomolny}
E.~B. Bogomolny.
\newblock {\em Sov. J. Nucl. Phys.}, {\bf{24}}:449, 1976.

\bibitem{Cieslinski}
Jan Cie\'{s}li\'{n}ski.
\newblock The darboux-bianchi-b\"{a}cklund transformation and soliton surfaces.
\newblock In J.~Cie\'{s}li\'{n}ski and D.~W\'{o}jcik, editors, {\em
  Nonlinearity $\&$ Geometry. Proceedings of First Non-Orthodox School},
  page~81. PWN, Warszawa, 1998.

\bibitem{Conte}
R.~Conte.
\newblock Exact solutions of nonlinear partial differential equations by
  singularity analysis.
\newblock In Antonio~M. Greco, editor, {\em Direct and Inverse Methods in
  Nonlinear Evolution Equations}, volume 632 of {\em Lecture Notes in Physics},
  page~1. Springer Verlag Berlin Heidelberg, 2003.

\bibitem{Debnath}
L.~Debnath.
\newblock {\em Nonlinear Partial Differential Equations for Scientists and
  Engineers}.
\newblock Springer Science+BusinessMedia, Birkh\"{a}user, 2012.

\bibitem{Castillo}
G.~F.~Torres del Castillo, C.~Andrade Miron, and R.~I.~Bravo Rojas.
\newblock {\em Riv. Mexic. Fis. E}, {\bf{59}}:140, 2019.

\bibitem{Doliwa}
A.~Doliwa.
\newblock Minimal surfaces, holomorphic curves and toda systems.
\newblock In J.~Cie\'{s}li\'{n}ski and D.~W\'{o}jcik, editors, {\em
  Nonlinearity \& Geometry. Proceedings of First Non-Orthodox School}, page
  227. PWN, Warszawa, 1998.

\bibitem{Elzaki}
T.~M. Elzaki and J.~Biazar.
\newblock {\em World Appl. Sci. J.}, {\bf{24}}:944, 2013.

\bibitem{Feng}
Z.~S. Feng.
\newblock {\em J. Phys. A: Math. Gen.}, {\bf{35}}:343, 2002.

\bibitem{Fushchich1993}
W.~I. Fushchich, W.~M. Shtelen, and N.~Serov.
\newblock {\em Symmetry Analysis and Exact Solutions of Equations of Nonlinear
  Mathematical Physics}.
\newblock Kluwer Academic Publishers, Dordrecht, 1993.

\bibitem{Gaeta}
G.~Gaeta.
\newblock {\em Nonlinear Symmetries and Nonlinear Equations}.
\newblock Springer Science+Business Media Dordrecht, 1994.

\bibitem{Goldstein}
P.~Goldstein.
\newblock Painleve test: standard gun of integrability hunter.
\newblock In J.~Cie\'{s}li\'{n}ski and D.~W\'{o}jcik, editors, {\em
  Nonlinearity $\&$ Geometry. Proceedings of First Non-Orthodox School}, page
  207. PWN, Warszawa, 1998.

\bibitem{Gu}
Ch. Gu, H.~Hu, and Z.~Zhou.
\newblock {\em Darboux Transformations in Integrable Systems. Theory and their
  Applications to Geometry}.
\newblock Springer, 2005.

\bibitem{Hosoya}
A.~Hosoya.
\newblock {\em Prog. Theor. Phys.}, {\bf{59}}:1781, 1978.

\bibitem{Krasilshchik}
I.~Krasil'shchik.
\newblock Symmetries and recursion operators for soliton equations.
\newblock In J.~Cie\'{s}li\'{n}ski and D.~W\'{o}jcik, editors, {\em
  Nonlinearity \& Geometry. Proceedings of First Non-Orthodox School}, page
  141. PWN, Warszawa, 1998.

\bibitem{Kudryashov}
N.~A. Kudryashov.
\newblock {\em Analytic Theory of Nonlinear Differential Equations}.
\newblock Moscow-Izhevsk, Institute of Computer Studies, 2004.
\newblock (in Russian).

\bibitem{Magri}
F.~Magri, G.~Falqui, and M.~Pedroni.
\newblock The method of poisson pairs in the theory of nonlinear pdes.
\newblock In Antonio~M. Greco, editor, {\em Direct and Inverse Methods in
  Nonlinear Evolution Equations}, volume 632 of {\em Lecture Notes in Physics},
  page~85. Springer Verlag Berlin Heidelberg, 2003.

\bibitem{Meleshko2005}
S.~V. Meleshko.
\newblock {\em Methods for Constructing Exact Solutions of Partial Differential
  Equations}.
\newblock Springer, 2005.

\bibitem{Musette}
M.~Musette.
\newblock Nonlinear superposition formulae of integrable partial differential
  equations by the singular manifold method.
\newblock In Antonio~M. Greco, editor, {\em Direct and Inverse Methods in
  Nonlinear Evolution Equations}, volume 632 of {\em Lecture Notes in Physics},
  page 137. Springer Verlag Berlin Heidelberg, 2003.

\bibitem{Nieszporski}
M.~Nieszporski and A.~Sym.
\newblock Weingarten congruences and non-auto-backlund transformations for
  hyperbolic surfaces.
\newblock In J.~Cie\'{s}li\'{n}ski and D.~W\'{o}jcik, editors, {\em
  Nonlinearity \& Geometry. Proceedings of First Non-Orthodox School}, page~37.
  PWN, Warszawa, 1998.

\bibitem{Olver}
P.~J. Olver.
\newblock {\em Applications of lie groups to differential equations}.
\newblock Springer-Verlag, 1986.

\bibitem{PolyaninZaitsev}
A.~D. Polyanin and V.~F. Zaitsev.
\newblock {\em Handbook of nonlinear partial differential equations}.
\newblock CRC Press Taylor \& Francis Group, 2012.

\bibitem{Rajchel}
K.~Rajchel and J.~Szcz\c{e}sny.
\newblock {\em Ann. Univ. Paedagog. Crac. Stud. Math.}, {\bf{XV}}:107, 2016.

\bibitem{Rubina}
L.~I. Rubina and O.~N. Ul'yanov.
\newblock {\em Vestn. Udmurtsk. Univ. Mat. Mekh. Komp. Nauki}, {\bf{27}}:355,
  2017.

\bibitem{Sanyal}
A.~K. Sanyal.
\newblock {\em Ann. Phys.}, {\bf{411}}:167971, 2019.
\newblock arXiv:1807.02769.

\bibitem{Satsuma}
J.~Satsuma.
\newblock Hirota bilinear method for nonlinear partial differential equations.
\newblock In Antonio~M. Greco, editor, {\em Direct and Inverse Methods in
  Nonlinear Evolution Equations}, volume 632 of {\em Lecture Notes in Physics},
  page 171. Springer Verlag Berlin Heidelberg, 2003.

\bibitem{SOK1}
K.~Sokalski.
\newblock {\em Acta Phys. Pol.}, {\bf{A56}}:571, 1979.

\bibitem{SOK1981}
K.~Sokalski.
\newblock {\em Phys. Lett.}, {\bf{A 81}}:102, 1981.

\bibitem{SOK1a}
K.~Sokalski.
\newblock {\em Acta Phys. Pol.}, {\bf{A 65}}:457, 1984.

\bibitem{SokStSok}
K.~Sokalski, {\L}.~St\c{e}pie\'{n}, and D.~Sokalska.
\newblock {\em J. Phys.}, {\bf{A35}}:6157, 2002.

\bibitem{SOK2}
K.~Sokalski, T.~Wietecha, and Z.~Lisowski.
\newblock {\em Acta Phys. Pol.}, {\bf{B32}}:17, 2001.

\bibitem{SOK2a}
K.~Sokalski, T.~Wietecha, and Z.~Lisowski.
\newblock {\em Acta Phys. Pol.}, {\bf{B32}}:2771, 2001.

\bibitem{SOK22}
K.~Sokalski, T.~Wietecha, and Z.~Lisowski.
\newblock {\em Int. J. Theor. Phys., Group Theory and Nonl. Optics, NOVA},
  {\bf{9}}:331, 2002.

\bibitem{SOK3}
K.~Sokalski, T.~Wietecha, and D.~Sokalska.
\newblock {\em J. Nonl. Math. Phys.}, {\bf{12}}:31, 2005.

\bibitem{ST1}
{\L}.~St\c{e}pie\'{n}.
\newblock {\em Bogomolny decomposition in the context of the concept of strong
  necessary conditions}.
\newblock {PhD} dissertation, Jagiellonian University, Krak\'{o}w, Poland,
  Marian Smoluchowski Institute of Physics, Department of Mathematics, Physics
  and Astronomy, Cracow, 2003.
\newblock in Polish.

\bibitem{SOK4}
{\L}.~St\c{e}pie\'{n}, D.~Sokalska, and K.~Sokalski.
\newblock {\em J. Nonl. Math. Phys.}, {\bf{16}}:25--34, 2009.

\bibitem{ST2010}
{\L}.~T. St\c{e}pie\'{n}.
\newblock {\em J. Comp Appl Math.}, {\bf{233}}:1607, 2010.

\bibitem{ST2}
{\L}.~T. St\c{e}pie\'{n}.
\newblock On bogomolny decompositions for the baby skyrme models.
\newblock In Piotr Kielanowski, S.~Twareque Ali, Alexander Odesskii, Anatol
  Odzijewicz, Martin Schlichenmaier, and Theodore Voronov, editors, {\em
  Geometric Methods in Physics. XXXI Workshop, Bia{\l}owie\.{z}a, Poland, June
  24-30, 2012}, Trends in Mathematics, pages 229--237. Birkh\"{a}user, Springer
  Basel, 2013.
\newblock math-ph/1204.6194 (published in April 2012).

\bibitem{ST2015}
{\L}.~T. St\c{e}pie\'{n}.
\newblock {\em Acta Phys. Pol. B}, {\bf{46}}:999, 2015.

\bibitem{ST11}
{\L}.~T. St\c{e}pie\'{n}.
\newblock {\em J. Phys. A}, {\bf{49}}:175202, 2016.

\bibitem{ST12}
{\L}.~T. St\c{e}pie\'{n}.
\newblock {\em J. Phys. A}, {\bf{51}}:015208, 2018.

\bibitem{ST3}
{\L}.~T. St\c{e}pie\'{n}.
\newblock {\em J. High Energ. Phys.}, {\bf{2020}}:140, 2020.

\bibitem{Szafirski}
B.~Szafirski.
\newblock private communication (2001).

\bibitem{Vitanov}
N.~K. Vitanov, Z.~I. Dimitrova, and K.~N. Vitanova.
\newblock {\em Appl. Math. Comp.}, {\bf{269}}:363, 2015.

\bibitem{WIET2005}
T.~Wietecha and K.~Sokalski.
\newblock {\em J. Symb. Comp.}, {\bf{44}}:1511, 2009.

\bibitem{Winternitz}
P.~Winternitz.
\newblock Lie groups, singularities and solutions of nonlinear partial
  differential equations.
\newblock In Antonio~M. Greco, editor, {\em Direct and Inverse Methods in
  Nonlinear Evolution Equations}, volume 632 of {\em Lecture Notes in Physics},
  page 223. Springer Verlag Berlin Heidelberg, 2003.

\bibitem{Zakharov1979}
V.~E. Zakharov and A.~B. Shabat.
\newblock {\em Funk. Anal. Pril.}, {\bf{13}}:13, 1979.

\bibitem{Zhang}
Z.-Y. Zhang, J.~Zhong, Sh.~Sh. Dou, J.~Liu, D.~Peng, and T.~Gao.
\newblock {\em Rom. Rep. Phys.}, {\bf{65}}:1155, 2013.

\end{thebibliography}

\end{document}